# Video Observation of Geminid 2010 from India using a portable system

Chintamani Pai[1], Ankush Bhaskar[2], Virendra Yadav[2], R. D. Tewari[3], A. M. Narsale[4]

[1]Department of Physics, University of Mumbai, India.

[2] Indian Institute of Geomagnetism, Navi Mumbai, India.

[3]Western Regional Instrumentation Centre, University of Mumbai, India.

[4] UM-DAE Centre for Excellence in Basic Sciences, Mumbai, India.

## Abstract:

Visual observations of meteor showers have been carried out for many decades by astronomers. Modern developments in imaging systems have advanced our knowledge about the shower dynamics and their origin. We have made an attempt to make a portable video recording system to observe meteor showers. The system consists of a camera capable of recording the rapid motion of meteors entering the Earth's atmosphere. The camera is interfaced with a laptop using an open source software like VirtualDub. The initial testing was carried out while observing Geminid meteor shower in December, 2010 from the base of Mahuli fort (Lat: 19.47°N, Long: 73.26°E) near Asangaon, India. Here, we present few of the meteors recorded during this event, followed by a preliminary analysis of the shower. The portable video recording system enables to capture meteors at a remote location. This system will strengthen traditional visual observation methods used by astronomers in India. Future coordinated studies with multi-station approach using such systems will assist in deriving the parameters associated with meteor shower activity and its impact on the Earth's atmosphere. Therefore, we propose long term and simultaneous multi-station video observations in the Indian subcontinent for continuous monitoring and better understanding of meteor dynamics.

# 1. Introduction:

Meteors are observed when a meteoroid or a speck of dust enters the Earth's atmosphere and ablates due to frictional forces resulting in a streak of light. Meteoroids are ranging from approximately 0.005 - 10 cm in diameter. They start vaporizing at an altitude ~ 120 km in the Earth's atmosphere lasting for few seconds to minutes. Observing a meteor is a random phenomenon considering the large area of the sky and time when it appears. However, during particular days of certain months meteors are observed frequently known as meteor showers. Several meteor observations methods have evolved over the years which include visual, photographic, radio and video observations [1].

Video observations of meteors have started in the 1970s using video tape recorders and TV monitors [2]. Japanese and Dutch observers were among the first to use this technique [3]. Later, these systems evolved when image intensifiers came into existence. Video observation of meteors is advantageous as compared to photographic techniques as it can detect duration and velocity of the meteors, which is not possible with photographic techniques [4-7]. Many video meteor observation networks exist in Europe due to international cooperation of video meteor observers, namely Networks like European Video Meteor Observation Network (*EDMONd*), network of Czech and Slovak observers known as Central European Meteor Network (*CEMeNt*), Polish Fireball Network (*PFN*), Hungarian Meteor Network (*HMN*), Italian Meteor Network (*IMTN*), French amateur network Base des Amateurs de Mtores (*BOAM*) etc [8-14]. International Meteor Organization (*IMO*), Germany started automated observations in 1997 establishing their video network. Since 2012, *IMO* Video Meteor Database (*IMO VMDB*) has been maintaining the records of single meteor observations over Europe connected to various networks. As of 2013, I*MO VMDB* has records of around

1,192,092 single meteor events [8-14]. Observers use software tools like *UFOCapture* or *MetRec* for multi-station observations of meteor events [1, 4, 15-16]. This helps in creating the orbits of meteoroids and determination of other relevant parameters. In Indian region, there is no such dedicated video network for monitoring meteor activity throughout the year. The present work is a first step towards making such a network reality in near future.

This paper describes the portable system developed by the authors for the video observation of meteors. Section 2 gives the details of the portable video recording system and observation technique. A preliminary analysis of few meteors captured using the system is also presented in Section 3. The papers is summarised in Section 4.

## 2. Observation system and Methods:

For observation of meteor showers, a clear night sky away from city is preferable. Light pollution free place enables observations of even faint meteors. Geminid meteor shower in December, 2010 was observed from a place known as Mahuli, Asangaon (Lat: 19.47ºN, Long: 73.26ºE) on the outskirts of the city of Mumbai, India. The location of the selected observation site is marked on the map (see Figure 1). The site provided a dark night sky with minimal light pollution. Visual and video observations were conducted for the shower from this location for approximately 7.5 hrs during the night of 14-15 December, 2010.

The meteor video recording system consists of a CCD camera (WAT-902 H2, 35.5 x 40 x 63 mm) with an optical lens attached to it, giving a field of view of approximately 90º. The major reason to choose this lens instead of a fish-eye lens was to avoid curvature effects while recording the meteors. WAT-902 H2 has a ½" CCD sensor with F1.4 aperture, capable of functioning in low light conditions with luminosity as low as 0.0001 lx. It requires an external DC power supply of 12 V. The output of the camera is coupled to a TV tuner card connected to a laptop for digitizing the output video signal (1 V pp, NTSC/PAL). This interface gives a frame rate of approximately 15 FPS. Open source software, VirtualDub, is used to capture the video stream from the camera. While recording the meteors, the camera was pointed upward (zenith) using a mini tripod. Video files were saved in .avi format. A block diagram of the portable meteor video system is shown in Figure 2.

The recorded videos were visually scanned for detection of meteors later on. The duration of the videos which had recorded meteor streaks were segmented and images in bmp/jpeg format were extracted using VirtulDub. Extracted images were further processed using open source image processing software ImageJ. Stacks of the images were created by using these image sequences. The flow chart of the meteor processing adopted in this study is shown in Figure 3.

## 3. Results:

As discussed in the previous section, images were extracted from the segmented videos showing meteor streaks. An example of the stacked images is shown in Figure 4. The duration and the direction of motion of the meteors are shown in each stacked frame. The duration of meteor streaks varied from 0.14 sec to 0.47 sec for the samples of meteor videos presented here. One can clearly see that generally the apparent magnitude of the meteors increased with time. There are background stars visible in some of the frames which can be utilized to get accurate apparent magnitude of the meteors. Further analysis of the meteor profiles will be carried out in near future.

Figure 5 shows visual observations of the Geminid meteor shower in 2010 conducted at the same site. Note that the shower activity peaks around 1-2 am IST on 15th December. The maximum hourly count of meteors reached ~ 40 meteors/hour, which is consistent with the observations reported by International Meteor Organization (http://www.imo.net/).

Histogram for the apparent magnitude of visually observed meteors is shown in Figure 6. It is seen that most of the visually observed meteors had an apparent magnitude of 2-3. However, a few bright meteors, having apparent magnitude < 0, were also observed. This distribution would imply the particle size/mass distribution of the meteor cloud through which the Earth passed on the day of the observations. The analysis of large number of recorded meteor videos will enable us to study the meteor cloud dynamics, evolution and the characteristics of each meteor and the cloud in detail.

## 4. Conclusions:

Video observation of meteors in low light conditions using a portable video system was successfully carried out. The system was used to observe Geminid meteor shower in December, 2010 from outskirts of Mumbai, India and promising preliminary results were obtained. Video observation of meteors can be valuable to study the comet and asteroid debris floating in space crossing the Earth's orbit. It is accurate and reliable as compared to traditional visual observations of meteors. Captured meteor data can be used for light curve analysis, monitoring of the shower activity, radiant determination, meteoroid size and mass distribution. Video observation data can be linked with radio observations of meteors revealing their dynamics in the Earth's atmosphere. Moreover, the data is useful for understanding and correlating ionospheric fluctuations as meteors enter the Earth's atmosphere. Therefore, realizing the importance of the continual monitoring of meteors, we propose to construct the Indian Meteor Network (IMN) adopting the presented or similar meteor video recording systems distributed across India.

**Acknowledgement:**

Authors thank Rahul Bhide, Jitendra Karekar, Neha Das, Mrityunjai Dabhade for participating in the meteor observation. Also, we extend our sincere thanks to WRIC staff, especially to Mr. Kutty for his technical support. Authors also thank Dr. Jaydeep Mukherjee, Director, FSGC, NASA, USA for valuable inputs and aspiration to pursue their interests.

**References:**

1. S. Molau, S. Nitschke, M. De Lignie, R. L. Hawkes, J. Rendtel, *Video Observations of Meteors: History, Current Status and Future Prospects*, WGN, the Journal of IMO **25**, 1 (1997).
2. Y. Fujiwara, *Television observations of meteors in Japan, Meteoroids and their parent bodies conference proceedings*, Slovac Academy of Sciences 265-268 (1993).
3. IMO Website, section on video observation system.
4. S. Molau, R. Ark, *Meteor shower radiant positions and structures as determined from single station video observations,* Planetary Space Science **45**, 857-864 (1997).
5. D. E. B. Fleming, R. L. Hawkes, J. Jones, *Light curves of faint television meteors*, Meteoroids and their parent bodies conference proceedings, Slovac Academy of Sciences 261-264 (1993).
6. Peter M. Millman, *One hundred and fifteen years of meteor spectroscopy*, International Astronomical Union conference proceedings (1980).
7. Milos Weber, *Results of ten years of photographic meteor spectroscopy*, WGN, the Journal of IMO **33**, 1 (2005).
8. Leonard Kornos, Jakub Koukal, Roman Piffl, Juraj Toth, *EDMOND Meteor Database*, Proceedings of the IMC, Poznań, 1 (2013).
9. Leonard Kornos, Jakub Koukal, Roman Piffl, Juraj Toth, *Database of meteoroid orbits from several European video networks*, Proceedings of the IMC, La Palma, 21 (2012).

10. J. M. Trigo-Rodriguez, A. J. Castro-Tirado, J. Lorca, J. Fabregat, V. J. Martinez, V. Reglero, M. Jelinek, P. Kubanek, T. Mateo, A. De Ugarte Postigo, *The development of the Spanish fireball network using a new all sky CCD system*, Earth, Moon, and Planets **95**, 553–567 (2004).
11. Sirko Molau, Jurgen Rendtel, *A Comprehensive List of Meteor Showers Obtained from 10 Years of Observations with the IMO Video Meteor Network*, WGN, the Journal of IMO **37**, 98 (2009).
12. S. Molau, M. Nitschke *"Computer based meteor search - a new dimension in video meteor observation"*, WGN 24, 119 (1996).
13. T. Sarma, J. Jones, *Double-station observations of 454 TV meteors*, Bull. Astron. Inst. Czech. **36**, 103 (1985).
14. Stanislav Vitek, Pavel Koten, Petr Pata, Karel Fliegel, *The double station automatic video observation of the meteors*, Advances in Astronomy, 943145 (2010).
15. S. Molau, *"MOVIE - Analysis of Video Meteors"*, Proceedings of the IMC 1994, 51 (1995).
16. Masao Kinoshita, Takuya Maruyama, Toru Sagayama, *Preliminary activity of Leonid meteor shower with a video camera in 1997* **26**, 41-44 (1999).

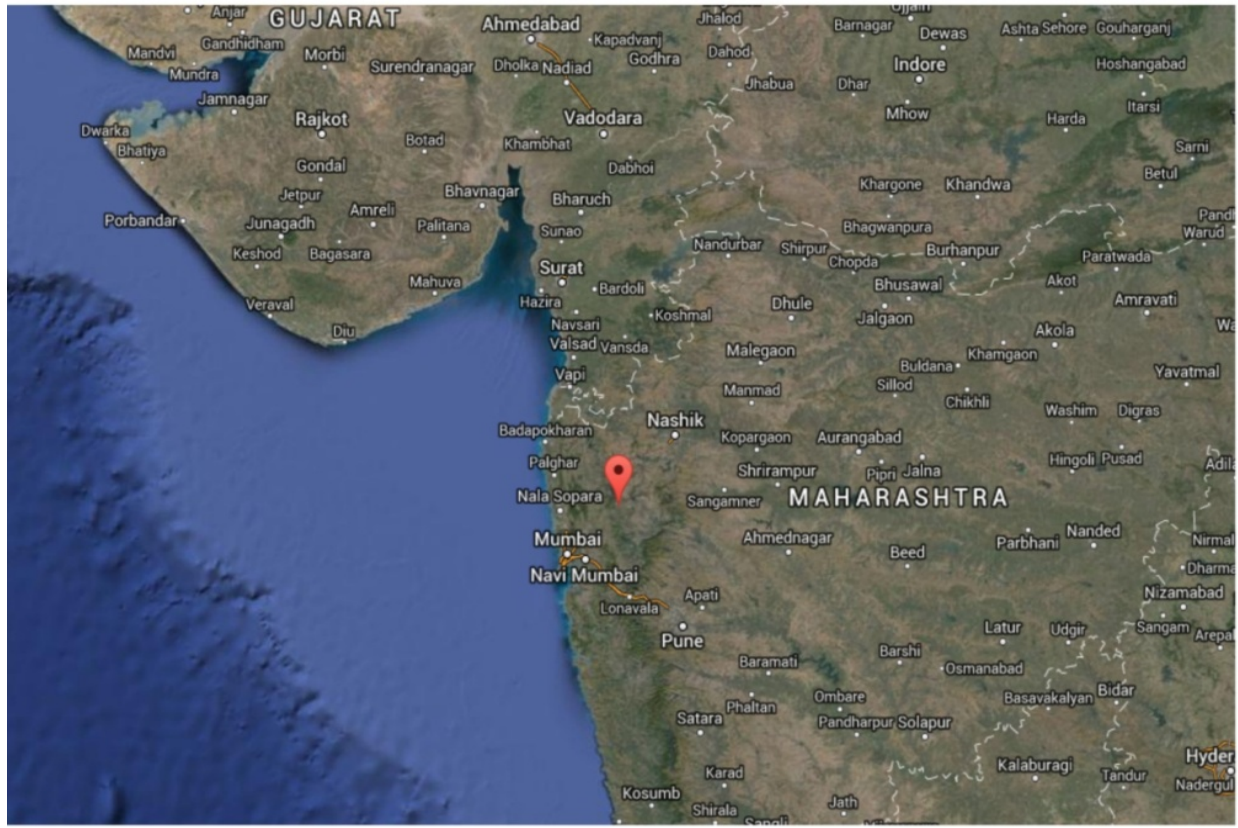

**Figure 1.** Map showing the location from where Geminid meteor shower, 2010 was observed (courtesy: Google Maps).

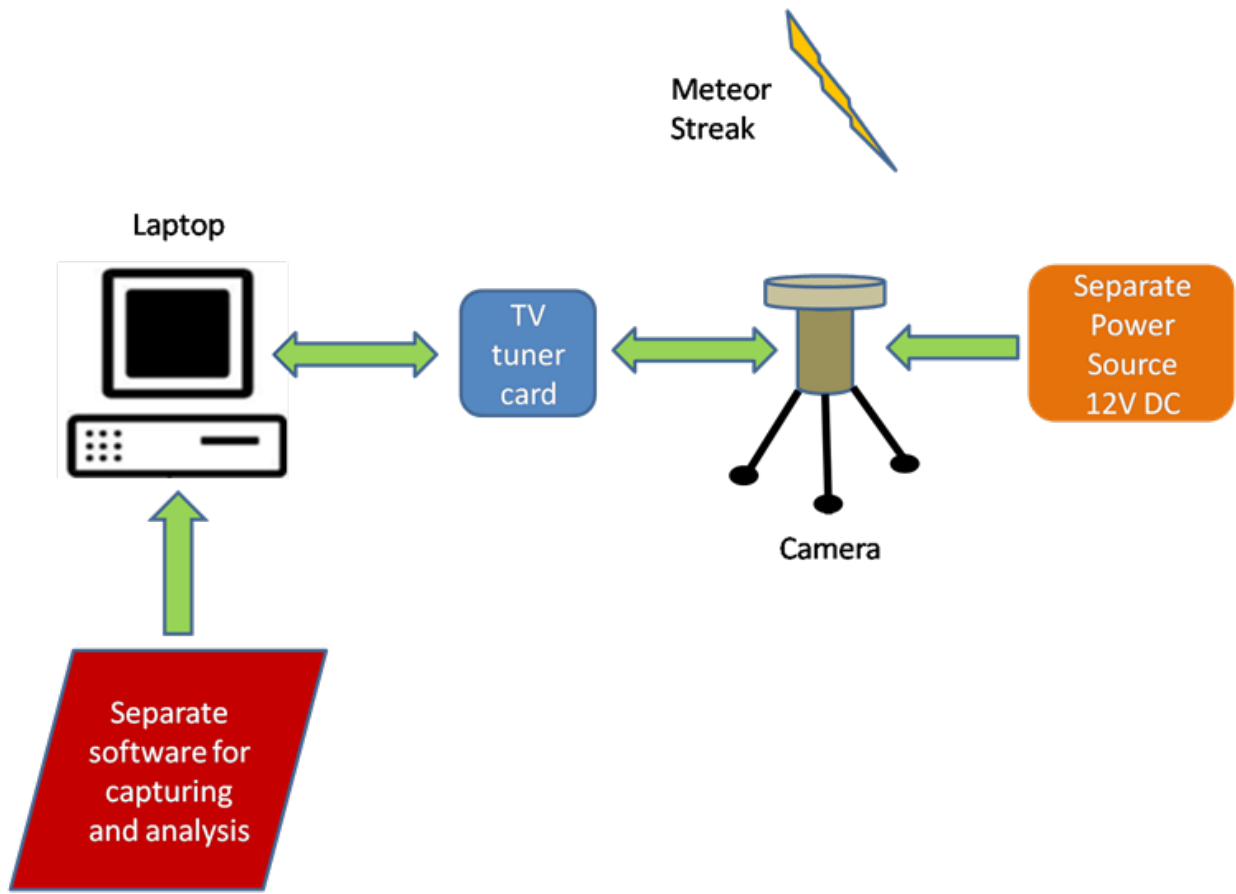

**Figure 2. Block diagram showing the experimental set-up of meteor video recording system.**

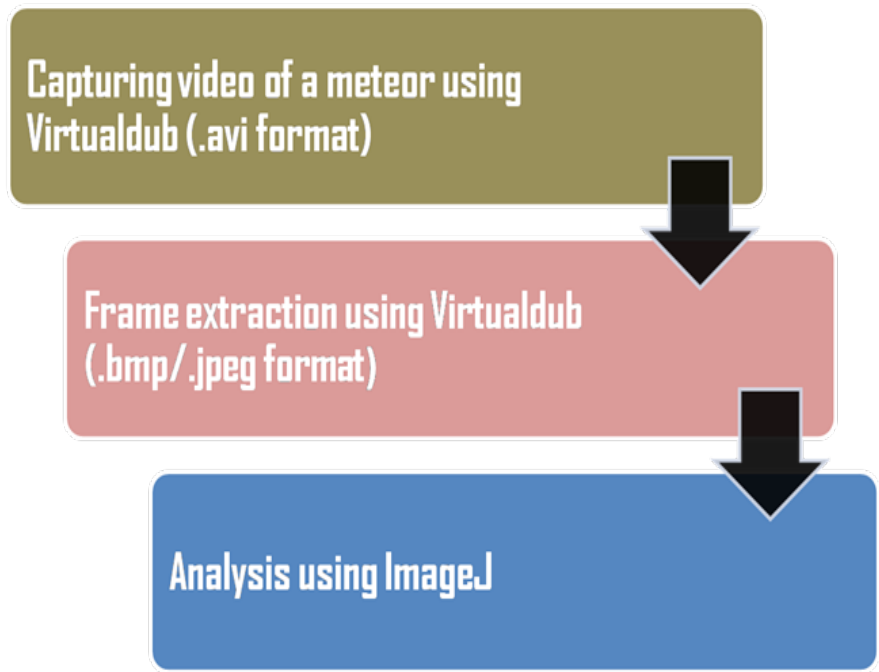

**Figure 3. Flow chart showing the step by step processing of the captured meteors by the video recording system.**

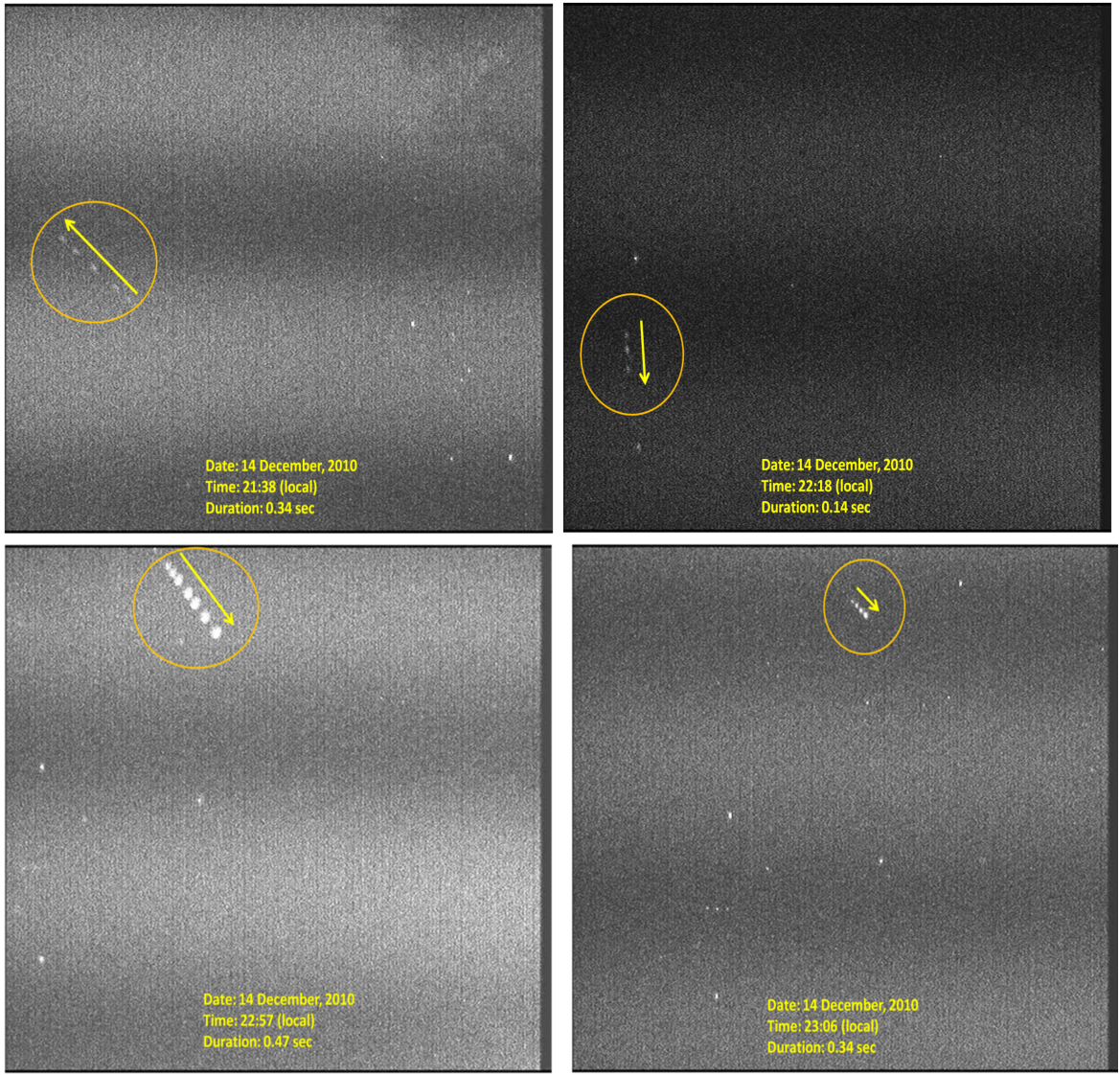

**Figure 4. Stacked frames for few meteors. Details are mentioned in each sub image. The meteors are encircled in yellow and the arrows point in the direction of the notion of meteors.**

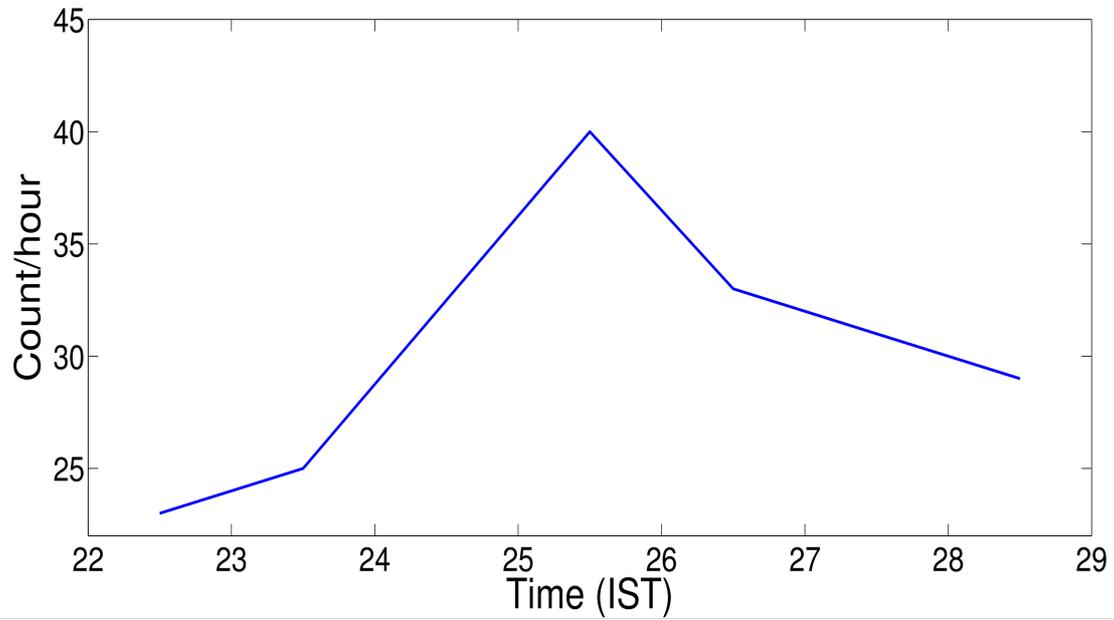

**Figure 5.** The observed activity of Geminid meteor shower from Mahuli, Asangaon, India during 14-15 December 2010.

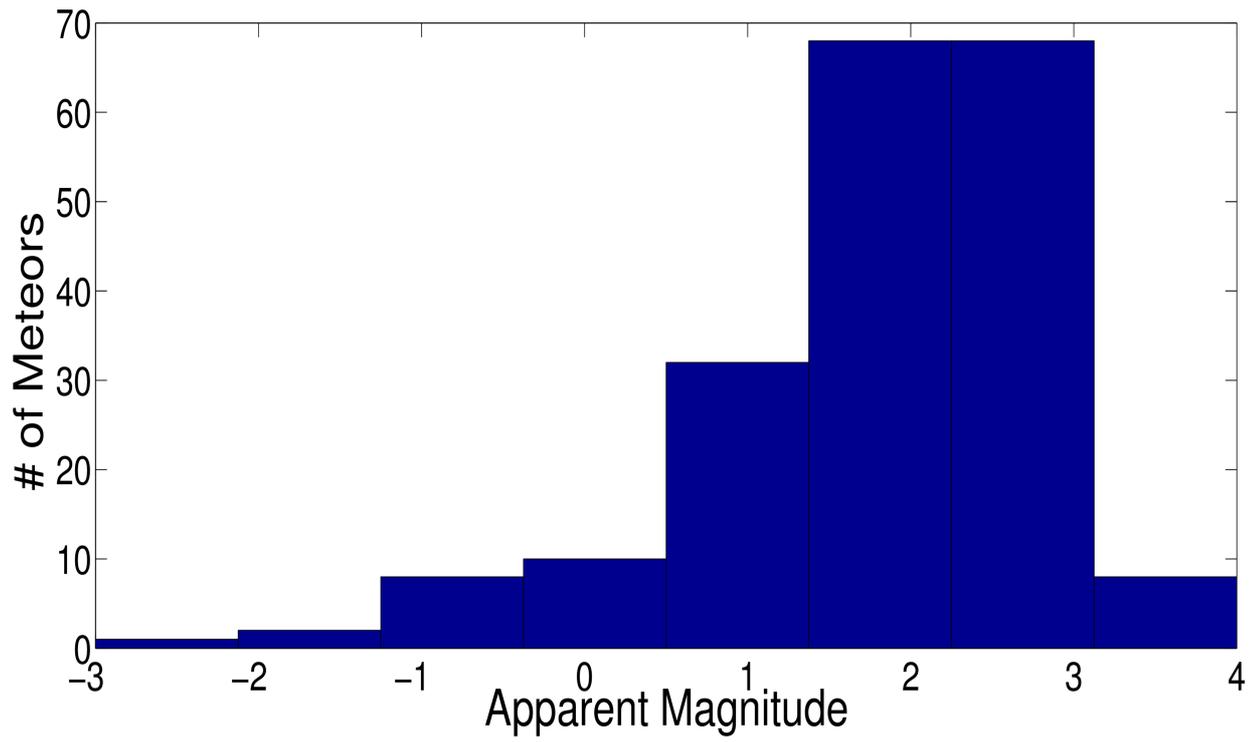

**Figure 6. Apparent magnitude distribution of the visually observed meteors of Geminid meteor shower during 14-15 December, 2010.**